\def\wP{\mbox{\large$\wp$}}
\def\Gn{\,\rule[-1mm]{.75mm}{4mm}\,}
\def\Z{|\hspace*{-.5mm}|\hspace*{-.5mm}|}
\def\wphi{\widehat{\varphi}^{\,(N)}}
\def\dphi{\dot{\widehat{\varphi}}\rule{0mm}{1ex}^{\,(N)}}
\def\ddphi{\ddot{\widehat{\varphi}}\rule{0mm}{1ex}^{\,(N)}}
\def\subs#1#2{\mbox{\small${#1\atop#2}$}}
\def\rf#1{(\ref{#1})}
\def\R{{\mathbb R}}
\def\T{{\mathbb T}}
\def\C{{\mathbb C}}
\def\e{{\rm e}}
\def\i{{\rm i}}
\documentclass{elsart}

\usepackage{amsmath,amssymb,graphicx}
\begin{document}
\begin{frontmatter}
\author[l1,l6]{R.~Chertovskih,\hspace*{-.8ex}}
\author[l3,l4,l5]{A.C.-L.~Chian,\hspace*{-.8ex}}
\author[l1]{O.~Podvigina,\hspace*{-.8ex}}
\author[l4,l5]{E.~Rempel,\hspace*{-.8ex}}
\author[l1]{V.~Zheligovsky}
\address[l1]{Institute of Earthquake Prediction Theory and Mathematical
Geophysics,\\Russian Academy of Sciences,\\
84/32 Profsoyuznaya St., 117997 Moscow, Russian Federation}
\address[l6]{Centre for Wind Energy and Atmospheric Flows,\\
Faculdade de Engenharia da Universidade do Porto,\\
Rua Dr.~Roberto Frias s/n, 4200-465 Porto, Portugal}
\address[l3]{Paris Observatory, LESIA, CNRS, 92195 Meudon, France}
\address[l4]{Institute of Aeronautical Technology (IEFM/ITA)\\
and World Institute for Space Environment Research (WISER),\\
S\~ao Jos\'e dos Campos, S\~ao Paulo 12228-900, Brazil}
\address[l5]{National Institute for Space Research (INPE),\\
P.O. Box 515, S\~ao Jos\'e dos Campos, S\~ao Paulo 12227-010, Brazil}
\title{Existence, uniqueness and analyticity\\of space-periodic solutions\\
to the regularised long-wave equation}
\begin{abstract}
We consider space-periodic evolutionary and travelling-wave solutions to the
regularised long-wave equation (RLWE) with damping and forcing. We establish
existence, uniqueness and smoothness of the evolutionary solutions for smooth
initial conditions, and global in time spatial analyticity of such solutions
for analytical initial conditions. The width of the analyticity strip decays
at most polynomially. We prove existence of travelling-wave solutions and
uniqueness of travelling waves of a sufficiently small norm. The importance
of damping is demonstrated by showing that the problem of finding travelling-wave
solutions to the undamped RLWE is not well-posed. Finally, we demonstrate
the asymptotic convergence of the power series expansion of travelling waves
for a weak forcing.
\end{abstract}
\end{frontmatter}

\section{Introduction}

The regularised long-wave equation (RLWE), also known as the
Benjamin--Bona--Mahony (BBM) equation, is a model for the propagation
of one-dimen\-sional, unidirectional small-amplitude long waves in
nonlinear dispersive media, being of great interest in the study of
propagation of long waves in shallow waters \cite{Dodd} such as tsunami
driven by an earthquake \cite{ToChiRem13} and drift waves in a controlled
nuclear fusion plasma \cite{HeSa89,Hort}. It was first derived
by Peregrine \cite{Pere}, then by Benjamin et al. \cite{BBM}, as an alternative
to the Korteweg-de Vries (KdV) equation \cite{Dodd}, in response to mathematical
difficulties associated with the KdV equation, such as the existence
and stability of solutions and other problems related to the dispersion term
\cite{BBM,ReMiChi09}. The RLWE was later derived by He \& Salat \cite{HeSa89}
as a model for nonlinear drift waves in plasmas, with a periodic driving
term and a linear damping term introduced ad hoc to study transition to chaos.

The understanding of the evolution of nonlinear physical systems such as
the RLWE requires a combined effort of numerical and analytical studies.
Numerically simulated nonlinear evolution of a driven-damped RLWE,
under the forcing of a periodic wave, has been analyzed in a series of papers.
He \& Chian \cite{HeChi03} discovered a new type of synchronization, the so-called
on-off collective imperfect phase synchronization, in the turbulent state
of RLWE solutions. In the driver frame, solutions to the RLWE can be represented
as a set of coupled oscillators in Fourier space. As the system evolves in time,
the oscillators in different spatial scales intermittently adjust
themselves to collective imperfect phase synchronization, inducing strong
bursts in the wave energy. Rempel \& Chian \cite{RemChi07} demonstrated
that non-attracting chaotic sets known as ``chaotic saddles'' are responsible
for transient and intermittent dynamics in the RLWE. As the driver amplitude is
increased, the system undergoes a transition from quasiperiodicity to temporal
chaos, then to spatiotemporal chaos. The resulting time series
in the spatiotemporal chaos regime display random switching between laminar
and bursty phases. Rempel \& Chian \cite{RemChi07} identified temporally
and spatiotemporally chaotic saddles which
are responsible for the laminar and bursty phases, respectively. Prior to the
transition to permanent spatiotemporal chaos, a spatiotemporally chaotic saddle
is responsible for chaotic transients that mimic the dynamics of the
post-transition attractor. Chian et al.~\cite{ChiMiRem10} applied the Fourier-Lyapunov
analysis to prove the duality of amplitude and phase synchronization in the RLWE due to
multiscale interactions in chaotic saddles at the onset of permanent
spatiotemporal chaos. By computing the power-phase spectral entropy and the
time-averaged power-phase spectra, they showed that the laminar/bursty states
in the on-off spatiotemporal intermittency correspond, respectively,
to the chaotic saddles with higher/lower degrees
of amplitude-phase synchronization across spatial scales.

From an analytical perspective, several works have presented studies on
the existence, uniqueness and stability of solutions to the RLWE.
In the seminal paper \cite{BBM}, Benjamin et al.~proved the existence
and uniqueness of nonperiodic solutions to the initial-value problem
for the RLWE in $\R^1$. While for the initial data in Sobolev spaces
$H^s(\R^1)$ for $s\ge0$ this problem for the RLWE is well-posed \cite{Bona09},
it is ill-posed for $s<0$ \cite{Pan11}. The stability of solitary-wave
solutions to the RLWE was shown by Bona \cite{Bona75}; existence and stability
of such solutions to the generalized BBM equation is examined in \cite{Zeng}
(see also references therein).
For the generalised RLWE with an arbitrary nonlinearity and a stronger damping
described by the Laplacian, space-periodic solutions have strong
finite-dimensional global attractors \cite{Wang97,WaYa} (see also
\cite{CeKaPo,SSW,Sta05}) in the Sobolev spaces $H^1(\T^1)$ and $H^2(\T^1)$;
the attractors consist of real analytical solutions \cite{Chu04}. Jafari et
al.~\cite{Jafa12} (see also\cite{EKNN}) found exact travelling-wave
solutions to the RLWE using the simplest equation method \cite{Kud1,Kud2}.

All the aforementioned papers examine the RLWE in its original form,
without the additional damping term introduced by He and Salat \cite{HeSa89}.
The goal of the present paper is to present the mathematical theory
of space-periodic solutions to the driven-damped RLWE. We begin by proving
existence, uniqueness (section \ref{exist}) and spatial analyticity (section
\ref{anal}) of space-periodic evolutionary solutions to the RLWE. In section
\ref{trwav} we show the existence of travelling-wave solutions to the damped
RLWE, as well as uniqueness of solutions whose norm does not exceed a certain
threshold (and hence a travelling-wave solution is unique,
provided the forcing is sufficiently week).
In section \ref{nonpos} we construct, in the form of infinite power series
in the inverse wave speed, a family of fast space-periodic travelling waves
that are formal asymptotic solutions to the zero-force RLWE without damping.
We do not prove convergence of these asymptotic power series; by construction,
upon truncation, the series represent travelling-wave solutions to the undamped
RLWE with some weak forcing, whose amplitude can be of the order
of any negative power of the wave speed. This shows that in the absence
of damping, finding travelling-wave solutions to the RLWE is not a well-posed
problem. The amplitude of forcing in numerical investigations
\cite{He,HeChi03,HeChi04,HeChi05,HeSa89,RemChi07,ReMiChi09,ToChiRem13}
of the RLWE was small. This has suggested to consider the asymptotic expansions
of solutions for a weak forcing; we do this in section \ref{wfor}.

\section{Existence and uniqueness of evolutionary solutions}\label{exist}

In this section we consider evolutionary solutions to the RLWE:
\begin{equation}
{\partial\over\partial t}(\varphi-a\varphi'')+b\varphi'+c\varphi\varphi'
+d\varphi+e(x,t)=0,
\label{rlwe}\end{equation}
where $'$ denotes differentiation in $x\in\R^1$, $a$, $b$, $c$ and $d$
are real constants, and $a>0$. The forcing $e(x,t)$ is prescribed.

Existence and uniqueness of the classical solutions to the forced BBM equation
(aka the non-damped RLWE, i.e., \rf{rlwe} for $d=0$) on the entire line $\R^1$
(the domain of the $x$ variable) was proved in \cite{BBM} under the assumption
that the initial ``energy'' $\int_{-\infty}^\infty(\varphi^2+(\varphi')^2)\,dx$
is finite, the forcing is continuous and has a finite Lebesgue norm.
By contrast, we consider solutions periodic in $x$ (assuming without any loss
of generality that the period is $2\pi$); existence and uniqueness of
space-periodic solutions to the BBM equation without forcing was proved (using
different techniques) in \cite{MM,Show} (see also \cite{MeMi}).

{\it Theorem 1.} Suppose $\varphi_0(x)\in C^\infty(\R^1)$ is $2\pi$-periodic and
$e(x,t)\in C^\infty(\R^1\times\R_+^1)$ is $2\pi$-periodic for any $t\ge0$.
For any constants $a>0,\,b,\,c$ and $d$ there exists a unique $2\pi$-periodic
solution to the RLWE, $\varphi(x,t)\in C^\infty(\R^1\times\R_+^1)$, such that
$\varphi(x,t)|_{t=0}=\varphi_0(x)$.

{\it Proof} exploits the general ideas involved in proofs of similar
statements for equations of the hydrodynamic type (see, e.g., \cite{Lad}).

$i.$ We use the Fourier-Galerkin method and consider an approximation
to the solution
$$\varphi^{(N)}(x,t)=\sum_n\wphi_n(t)\,\e^{\i nx},$$
where $\wphi_n=0$ for $|n|>N$.
The approximate Fourier coefficients $\wphi_n(t)$ satisfy the equations
obtained by the orthogonal projection in $L_2([0,2\pi])$ of the RLWE onto
the subspace spanned by the Fourier harmonics $\e^{\i nx}$ for $|n|\le N$:
\begin{align}
&(1+an^2)\dphi_n+(d+\i bn)\wphi_n+\i c\sum_mm\wphi_m\wphi_{n-m}+\widehat{e}_n(t)=0,\label{geq}\\
&\left.\rule[-1.5ex]{0mm}{1em}\wphi_n(t)\right|_{t=0}=\widehat{\varphi}_{0,n},\nonumber
\end{align}
where the dot denotes differentiation in time, and $\widehat{e}_n(t)$ and $\widehat{\varphi}_{0,n}$
are the Fourier coefficients of $e(x,t)$ and $\varphi_0(x)$, respectively:
$$e(x,t)=\sum_n\widehat{e}_n(t)\e^{\i nx},\qquad
\varphi_0(x)=\sum_n\widehat{\varphi}_{0,n}\e^{\i nx}.$$

We employ the seminorms $\|\cdot\|_s$ defined as follows:
for $f(x)=\sum_n\widehat{f}_n\e^{\i nx}$,
$$\|f\|^2_s=\left\{\begin{array}{lr}
\displaystyle\sum_n|\widehat{f}_n|^2|n|^{2s},&s>0;\\
\displaystyle\sum_n|\widehat{f}_n|^2(\max(|n|,1))^{2s},~\rule{0mm}{3ex}&s\le0.
\end{array}\right.$$
For $s\ge 0$, $\|\cdot\|^2_0+\|\cdot\|^2_s$ is the square of the norm
in the Sobolev space $H^s(\T^1)$ of $2\pi$-periodic functions.

$ii.$ An energy bound, on which all our constructions are based, is obtained
by multiplying \rf{geq} by $\wphi_{-n}=\overline{\wphi_n}$ and
summing the results over all $n$:
\begin{equation}
{1\over 2}\,{d\over dt}\left(\|\varphi^{(N)}\|_0^2+a\|\varphi^{(N)}\|_1^2\right)
+d\|\varphi^{(N)}\|_0^2=-\sum_n\wphi_{-n}\widehat{e}_n(t).
\label{enb}\end{equation}
The sums involving constants $b$ and $c$ vanish, since by periodicity
$$\i b\sum_nn\wphi_n\wphi_{-n}
={b\over2\pi}\int_0^{2\pi}\varphi^{(N)}{d\over dx}\varphi^{(N)}\,dx
={b\over4\pi}\int_0^{2\pi}{d\over dx}(\varphi^{(N)})^2\,dx=0$$
and
$$\i c\sum_{m,n}m\wphi_m\wphi_{n-m}\wphi_{-n}
={c\over2\pi}\int_0^{2\pi}(\varphi^{(N)})^2{d\over dx}\varphi^{(N)}\,dx
={c\over6\pi}\int_0^{2\pi}{d\over dx}(\varphi^{(N)})^3\,dx=0.$$
%{\small This can be also proved in the language of sums, for instance the last
%identity is obtained as follows:
%$$\i c\sum_{m,n}m\wphi_m\wphi_{n-m}\wphi_{-n}={\i c\over3}
%\left(\sum_{m,n}m\wphi_m\wphi_{n-m}\wphi_{-n}
%+\sum_{m,n}m\wphi_m\wphi_{n-m}\wphi_{-n}
%+\sum_{m,n}m\wphi_m\wphi_{n-m}\wphi_{-n}\right);$$
%change here the indices $m\to-n$, $n\to m-n$ in the second sum and
%$m\to n-m$ in the third sum and continue the transformation:
%$$={\i c\over3}
%\left(\sum_{m,n}m\wphi_m\wphi_{n-m}\wphi_{-n}
%+\sum_{m,n}(-n)\wphi_{-n}\wphi_m\wphi_{n-m}
%+\sum_{m,n}(n-m)\wphi_{n-m}\wphi_m\wphi_{-n}\right)$$
%$$={\i c\over3}
%\sum_{m,n}(m-n+(n-m))\,\wphi_m\wphi_{n-m}\wphi_{-n}=0.$$}

By Gronwall's lemma, identity \rf{enb} implies the inequality
$$\left(\|\varphi^{(N)}\|_0^2+a\|\varphi^{(N)}\|_1^2\right)^{1/2}
\le C_0(t)$$
(recall that $a>0$), where
$$C_0(t)\equiv\left(\|\varphi_0\|_0^2+a\|\varphi_0\|_1^2\right)^{1/2}\e^{\widetilde{d}t}
+\int_0^t\|e(x,\tau)\|_0\,\e^{\widetilde{d}(t-\tau)}\,d\tau,$$
$$\widetilde{d}=\left\{
\begin{array}{lr}
0,&d\ge 0,\\
|d|,\rule{0mm}{3ex}&d<0.
\end{array}\right.$$
From this inequality we infer bounds, that are uniform in $N$:
$\|\varphi^{(N)}\|_s\le C_s(t)$ for $s=0$ and 1 (we can set
$C_1(t)=C_0(t)/\sqrt{a}$).

$iii$ We derive now bounds, that are uniform in $N$, for $\|\varphi^{(N)}\|_s$,
where $s>0$ is arbitrary.

For $s>1$, multiply \rf{geq} by $\wphi_{-n}|n|^{2s}(1+an^2)^{-1}$ and sum
the results over $n$:
\begin{align}%\hspace*{-1em}
{1\over 2}\,{d\over dt}\|\varphi^{(N)}\|_s^2
=&-\sum_n{(d+\i bn)|n|^{2s}\over1+an^2}\wphi_n\wphi_{-n}
-\i c\sum_{m,n}{m|n|^{2s}\over1+an^2}\wphi_m\wphi_{n-m}\wphi_{-n}\nonumber\\
&-\sum_n{|n|^{2s}\over1+an^2}\wphi_{-n}\widehat{e}_n(t).\label{enbp}
\end{align}
We bound each sum in the r.h.s. By the Cauchy-Bunyakovsky-Schwarz inequality,
$$-\sum_n{d|n|^{2s}\over1+an^2}\,\wphi_n\wphi_{-n}\le
{\widetilde{d}\over a}\,\|\varphi^{(N)}\|_s\|\varphi^{(N)}\|_{s-2}.$$
By changing the index of summation $n\to -n$, we establish
$$\sum_n{\i bn|n|^{2s}\over1+an^2}\,\wphi_n\wphi_{-n}=0.$$
To bound the third sum, note that, by the same inequality,
\begin{equation}
\sum_n{|n|^{s+1}\over1+an^2}\,|\wphi_{-n}|\le Q_1\|\varphi^{(N)}\|_s
\label{Linfb}\end{equation}
$$\mbox{for~~}Q_1=\left(2\sum_{n>0}\left({n\over1+an^2}\right)^2\right)^{1/2}
\mbox{~and any~}s\ge0,$$
and
\begin{equation}
|m|\,|n|^{s-1}\le Q_{2,s}(|m|^s+|n-m|^s)
\label{mnp}\end{equation}
\noindent
for all $m$, $n$, $s\ge1$ and some suitable constants $Q_{2,s}$. Therefore,
\begin{align*}
&\left|c\sum_{m,n}{m|n|^{2s}\over1+an^2}\wphi_m\wphi_{n-m}\wphi_{-n}\right|\\
\le&|c|\sum_{n}\left(\sum_{m}Q_{2,s}(|m|^s+|n-m|^s)|\wphi_m||\wphi_{n-m}|\right)
{|n|^{s+1}\over1+an^2}\,|\wphi_{-n}|\rule[-4ex]{0mm}{6ex}\\
\le&2|c|Q_{2,s}\|\varphi^{(N)}\|_s\|\varphi^{(N)}\|_0\,Q_1\|\varphi^{(N)}\|_s
=Q_{3,s}\|\varphi^{(N)}\|_0\|\varphi^{(N)}\|^2_s,
\end{align*}
where $Q_{3,s}=2|c|Q_1Q_{2,s}$. Finally,
$$\left|\sum_n{|n|^{2s}\over1+an^2}\wphi_{-n}\widehat{e}_n(t)\right|\le
{1\over a}\,\|\varphi^{(N)}\|_s\|e(x,t)\|_{s-2}.$$
Collecting all the bounds, we obtain from \rf{enbp}:
$${1\over 2}\,{d\over dt}\|\varphi^{(N)}\|_s^2\le
{\widetilde{d}\over a}\,\|\varphi^{(N)}\|_s\|\varphi^{(N)}\|_{s-2}
+Q_{3,s}\|\varphi^{(N)}\|_0\|\varphi^{(N)}\|^2_s
+{1\over a}\,\|\varphi^{(N)}\|_s\|e(x,t)\|_{s-2},$$
i.e.
$${d\over dt}\|\varphi^{(N)}\|_s\le{\widetilde{d}\over a}\,\|\varphi^{(N)}\|_{s-2}
+Q_{3,s}\|\varphi^{(N)}\|_0\|\varphi^{(N)}\|_s+{1\over a}\,\|e(x,t)\|_{s-2}.$$
Using Gronwall's lemma, we deduce by induction from this inequality bounds,
that are uniform in~$N$:
\begin{align}
\|\varphi^{(N)}(x,t)\|_s\le&
\|\varphi^{(N)}(x,0)\|_s\,\e^{-Q_{3,s}\int_0^tC_0(\tau)\,d\tau}\nonumber\\
&+\,{1\over a}\int_0^t\left(\widetilde{d}C_{s-2}(\tau)+\|e(x,\tau)\|_{s-2}\right)
\,\e^{-Q_{3,s}\int_\tau^t C_0(\tau')\,d\tau'}\,d\tau
\label{phNbou}\end{align}
for all even $s\ge2$. We denote the r.h.s.~of \rf{phNbou} by $C_s(t)$.
By interpolation,
$\|\varphi^{(N)}(x,t)\|_s\le C_s(t)\equiv C^{1-\mu}_{S+2}(t)\,C^{\mu}_S(t)$ holds true
for any $s\ge0$, where $S\ge0$ is integer, $0\le\mu\le 1$ and $s=(1-\mu)(S+2)+\mu S$.
(The specific form of the bounding functions $C_s(t)$ is not important for our purposes.)

$iv.$ We derive now bounds for $\|\dot{\varphi}^{(N)}\|_s$
that are uniform in $N$.

Multiply \rf{geq} by $\dphi_{-n}|n|^{2s}(1+an^2)^{-1}$ and sum
the results over $n$:
\begin{align}
\|\dot{\varphi}^{(N)}\|_s^2
=&-\sum_n{(d+\i bn)|n|^{2s}\over1+an^2}\wphi_n\dphi_{-n}
-\i c\sum_{m,n}{m|n|^{2s}\over1+an^2}\wphi_m\wphi_{n-m}\dphi_{-n}\nonumber\\
&-\sum_n{|n|^{2s}\over1+an^2}\dphi_{-n}\widehat{e}_n(t).\label{enbpdot}
\end{align}
We derive bounds for each sum in the r.h.s. for $s\ge0$. Clearly,
$$\left|\sum_n{(d+\i bn)|n|^{2s}\over1+an^2}\,\wphi_n\dphi_{-n}\right|\le
Q_4\,\|\dot{\varphi}^{(N)}\|_s\|\varphi^{(N)}\|_{s-1},$$
where $Q_4$ is a constant such that $Q_4(1+an^2)\ge\max(1,|n|)(|d|+|b||n|)$
for all~$n$. For $s=0$, the second sum can be bounded as follows:
$$\left|c\sum_{m,n}{m\over1+an^2}\wphi_m\wphi_{n-m}\dphi_{-n}\right|\le
Q_{3,0}\|\varphi^{(N)}\|_0\|\varphi^{(N)}\|_1\|\dot{\varphi}^{(N)}\|_0,$$
where
$$Q_{3,0}=|c|\left(\sum_n(1+an^2)^{-2}\right)^{1/2}.$$
For $s\ge1$, we use inequalities \rf{Linfb} applied
to $\dot{\varphi}^{(N)}$ instead of $\varphi^{(N)}$ and \rf{mnp}:
\begin{align*}
&\left|c\sum_{m,n}{m|n|^{2s}\over1+an^2}\wphi_m\wphi_{n-m}\dphi_{-n}\right|\\
\le&|c|\sum_{n}\left(\sum_{m}Q_{2,s}(|m|^s+|n-m|^s)|\wphi_m||\wphi_{n-m}|\right)
{|n|^{s+1}\over1+an^2}\,|\dphi_{-n}|\rule[-4ex]{0mm}{6ex}\\
\le&2|c|Q_{2,s}\|\varphi^{(N)}\|_s\|\varphi^{(N)}\|_0\,Q_1\|\dot{\varphi}^{(N)}\|_s
=Q_{3,s}\|\varphi^{(N)}\|_0\|\varphi^{(N)}\|_s\|\dot{\varphi}^{(N)}\|_s.
\end{align*}
Finally, for $s\ge0$ the last sum satisfies the inequality
$$\left|\sum_n{|n|^{2s}\over1+an^2}\dphi_{-n}\widehat{e}_n(t)\right|\le
\max(1,a^{-1})\,\|\dot{\varphi}^{(N)}\|_s\|e(x,t)\|_{s-2}.$$
Collecting the bounds, we obtain for $s=0$ and $s\ge 1$ from \rf{enbpdot}:
\begin{equation}
\|\dot{\varphi}^{(N)}\|_s\le Q_4\,\|\varphi^{(N)}\|_{s-1}+
Q_{3,s}\|\varphi^{(N)}\|_0\|\varphi^{(N)}\|_{\max(s,1)}
+\max(1,a^{-1})\,\|e(x,t)\|_{s-2}.
\label{Dsdef}\end{equation}
By induction, \rf{Dsdef} yields a bound that is uniform in $N$, for any integer
$s\ge0$. We denote the r.h.s.~of \ref{Dsdef} by $D_s(t)$. By interpolation,
for any $s\ge0$
\begin{equation}
\|\dot{\varphi}^{(N)}(x,t)\|_s\le D^{1-\mu}_{S+1}(t)D^{\mu}_S(t)\equiv D_s(t)
\label{dNbou}\end{equation}
at any time $t\ge0$, where $S\ge0$ is integer, $0\le\mu\le 1$ and $s=(1-\mu)(S+1)+\mu S$.

$v.$ Differentiating \rf{geq} in time, we find
$$(1+an^2)\ddphi_n+(d+\i bn)\dphi_n+\i c\sum_m
m(\dphi_m\wphi_{n-m}+\wphi_m\dphi_{n-m})+\dot{e}_n(t)=0.$$
Using this equation and the bounds for $\|\varphi^{(N)}\|_s$ and
$\|\dot{\varphi}^{(N)}\|_s$ obtained above for arbitrarily large $s$,
it is easy to show that $|\ddphi_n|$ are uniformly bounded in $N$
for each $n$.

Consider a time interval $[0,T]$ for some $T>0$. We have demonstrated that,
for each $n$, $|\dphi_n(t)|$ and $|\ddphi_n(t)|$
are uniformly bounded in $N$ and $t\in[0,T]$, and hence the sets of functions
$\wphi_n(t)$ and $\dphi_n(t)$ are equicontinuous.
Therefore, applying the Arzel\`a--Ascoli theorem and using the diagonal process,
we can choose a subsequence $N_k\to\infty$ such that\\
1) for each $n$, $\varphi^{(N_k)}_n$ and $\dot{\varphi}^{(N_k)}_n$ uniformly
converge to some continuous functions $\varphi_n$ and $\delta_n$; furthermore,
$\delta_n=\dot{\varphi}_n$, as can be seen by letting $N_k\to\infty$ in the relation
$$\wphi_n(t)=\varphi_n(0)+\int_0^t\dphi_n(\tau)\,d\tau;$$
2) the bounds
\begin{equation}
\|\varphi(x,t)\|_s\le C_s(t)
\label{phbou}\end{equation}
and
\begin{equation}
\|\dot{\varphi}(x,t)\|_s\le D_s(t)
\label{dbou}\end{equation}
hold true for the limit functions
$$\varphi(x,t)=\sum_n\varphi_n(t)\e^{\i nx},\qquad
\dot{\varphi}(x,t)=\sum_n\dot{\varphi}_n(t)\e^{\i nx}$$
(this can be shown by considering inequalities
\rf{phNbou} and \rf{dNbou} for $N_k\to\infty$).

Thus, at each time $t$ the limit functions $\varphi(x,t)$ and
$\dot{\varphi}(x,t)$ are infinitely smooth in $x$ (provided the initial data
and the forcing are infinitely smooth). In the limit
$N_k\to\infty$, the Galerkin equation \rf{geq} becomes
\begin{equation}
(1+an^2)\dot{\varphi}_n+(d+\i bn)\varphi_n+\i c\sum_m
m\varphi_m\varphi_{n-m}+\widehat{e}_n(t)=0
\label{Feq}\end{equation}
(passing to the limit in the infinite sum in $m$ is possible, because
the sum converges uniformly in $N$). Relations \rf{Feq} imply that
$\varphi(x,t)$ satisfies the original RLWE in the classical sense.

Differentiating the RLWE $s-1$ times in $t$, we incrementally establish
(by induction in $s$) that \hbox{${\partial^s\over\partial t^s}
(\varphi-a\varphi'')$} and hence $\partial^s\varphi/\partial t^s$ are continuous
in time; this proves that $\varphi(x,t)\in C^\infty(\R^1\times\R_+^1)$.

Finally, if there exist two distinct smooth solutions to the RLWE, application
of Gronwall's lemma to the linear equation for the difference between them
establishes that the difference is zero. In particular, the limit
functions obtained for different subsequences $N_k\to\infty$ and/or on
different time intervals $[0,T]$ necessarily coincide. Q.E.D.

\section{Spatial analyticity of evolutionary solutions}\label{anal}

Temporal analyticity of solutions to the zero-force BBM equation was proved
in \cite{BBM}. These authors analysed convergence of Taylor's expansion
of the solution in time, employing an integral operator that maps the $m$-th
time derivative of the solution to the time derivative of order $m+1$. Here we
prove the spatial analyticity of $\varphi$ by the techniques of \cite{Zh}.

For any $\sigma>0$ we define the Gevrey--Sobolev seminorms
of $f(x)=\sum_n\hat{f}_n\e^{\i nx}$ by the relation
$$\Gn f\Gn^2_{\sigma,s}=\left\{\begin{array}{lr}
\displaystyle\sum_n|\widehat{f}_n|^2\,|n|^{2s}\,\e^{2\sigma|n|},&s>0;\\
\displaystyle\sum_n|\widehat{f}_n|^2\,(\max(|n|,1))^{2s}\,\e^{2\sigma|n|},~\rule{0mm}{3ex}&s\le0.
\end{array}\right.$$
Functions, whose Gevrey--Sobolev norms are finite, are analytic;
the first index $\sigma$ is a lower estimate of the width
of the analyticity strip of $f$ around the real axis on the complex plane.

We also introduce a seminorm
$$\Z f\Z^2=\sum_n(|n|+a|n|^3)|\widehat{f}_n|^2$$
equivalent to $\|\cdot\|_{3/2}$.

{\it Theorem 2.} Suppose $\varphi_0(x)$ and $e(x,t)$ satisfy the conditions
of Theorem 1 and are analytic in~$x$: for some constants $\sigma>0$ and
$\beta>0$, $\Gn\varphi_0(x)\Gn_{\sigma,\,3/2}<\infty$ and
$\Gn e(x,t)\Gn_{\beta,\,0}$ is uniformly bounded in time. Then the solution
to the RLWE is analytic in $x$ at any $t\ge0$, and the width of its analyticity
strip around the real axis decreases in time at most exponentially. For $d\ge0$,
the width decreases in time at most algebraically.

{\it Proof}.
We will show that at any time $t$ the solutions to the Fourier--Galerkin system
of equations \rf{geq}, that were considered in Theorem 1, for some $\kappa(t)>0$
have Gevrey--Sobolev norms $\Gn\varphi^{(N)}(x,t)\Gn_{\kappa(t),\,3/2}$,
that are bounded uniformly in $N$. This will imply that the solution to the RLWE,
$\varphi(x,t)$, also have finite norms $\Gn\varphi(x,t)\Gn_{\kappa(t),\,3/2}$,
this proving Theorem 2.

For a given $N$, we consider a transformation
\begin{equation}
\wphi_n(t)=\widehat{w}^{(N)}_n(t)\exp\left(-{\beta|n|\over
1+\Z w^{(N)}(x,t)\Z^{1+\varepsilon}}\right),
\label{subs}\end{equation}
$$w^{(N)}(x,t)=\sum_n\widehat{w}^{(N)}_n(t)\,\e^{\i nx},$$
where $\varepsilon\le 1$ is a positive constant. For brevity,
we henceforth omit the superscript $(N)$ in $\widehat{w}^{(N)}_n$. We seek
a solution to the system of nonlinear equations \rf{subs} in the form
$$\widehat{w}_n(t)=\widehat{\varphi}_n(t)\exp(\psi(t)|n|),$$
where $\psi(t)>0$ satisfies the equation
$$\psi(t)\left(1+\left(\sum_{|n|\le N}(|n|+a|n|^3)\,
\e^{2\psi(t)|n|}\,|\wphi_n(t)|^2\right)^{\!\!(1+\varepsilon)/2}\,\right)=\beta.$$
It has a unique solution for any $t\ge0$, because the l.h.s.~is a continuous
monotonically increasing unbounded function of $\psi$, that vanishes for
$\psi=0$. We assume without any loss of generality that
$$\beta\le\sigma\left(1+(\Gn\varphi_0\Gn_{\sigma,\,1/2}^2
+a\Gn\varphi_0\Gn_{\sigma,\,3/2}^2)^{(1+\varepsilon)/2}\,\right),$$
whereby $\|w^{(N)}(x,t)\|_{3/2}$ are bounded uniformly in $N$ at $t=0$.

Substitution \rf{subs} transforms the Fourier--Galerkin equations \rf{geq}
into the system of equations
$$(1+an^2)\dot{\widehat{w}}_n+\beta(1+\varepsilon)(|n|+a|n|^3)\,\widehat{w}_n\,
{\Z w\Z^{\varepsilon}\over(1+\Z w\Z^{1+\varepsilon})^2}
{d\over dt}\Z w\Z+(d+\i bn)\widehat{w}_n$$
$$+\,\i c\sum_mm\widehat{w}_m\widehat{w}_{n-m}
\e^{\gamma(|n|-|{k}|-|{n-k}|)}+\widehat{e}_n(t)\e^{\gamma|n|}=0,$$
where it is denoted $\gamma=\beta\,/\,(1+\Z w\Z^{1+\varepsilon})$.

Multiplying the equation by $w_{-n}=\overline{\widehat{w}_n}$
and summing up the results over $n$, we find
\begin{align}
&{1\over2}\,{d\over dt}\left(\|w\|^2_0+a\|w\|^2_1
+2\beta(1+\varepsilon)I(\Z w\Z)\right)+d\|w\|^2_0\nonumber\\
&+\,\i c\sum_{m,n}m\widehat{w}_m\widehat{w}_{n-m}\widehat{w}_{-n}
\e^{\gamma(|n|-|m|-|n-m|)}+\sum_n\widehat{e}_n(t)\e^{\gamma|n|}\widehat{w}_{-n}=0,\rule{0mm}{4ex}\label{enin}
\end{align}
where it is denoted
$$I(q)=\int_0^q{u^{2+\varepsilon}\over(1+u^{1+\varepsilon})^2}\,du.$$
For $0<\varepsilon<1$ and large $q$,
\begin{equation}
I(q)=(1-\varepsilon)^{-1}\,q^{1-\varepsilon}+O(q^{-2\varepsilon}).
\label{ord}\end{equation}

We transform now the sum
$$\sum\equiv\i c\sum_{m,n}m\widehat{w}_m\widehat{w}_{n-m}\widehat{w}_{-n}
\,\e^{\gamma(|n|-|m|-|n-m|)}.$$
It remains unaltered when we change the index $m\to n-m$, as well as when
we change the indices $m\to-n$, $n\to m-n$. Summing the two sums obtained
by these changes of indices with the original sum, we find
$$\sum={\i c\over3}\sum_{m,n}n\widehat{w}_m\widehat{w}_{n-m}\widehat{w}_{-n}
\,(\e^{\gamma(|n|-|m|-|n-m|)}-\e^{\gamma(|n-m|-|n|-|m|)}).$$
By virtue of the inequalities $|\e^{\mu'}-\e^{\mu''}|\le|\mu'-\mu''|$ that
holds true for any $\mu'\le 0$ and $\mu''\le0$, and
$|n|^\mu\le|m|^\mu+|n-m|^\mu$ for any $0\le\mu\le1$, the above relation implies
\begin{equation}
\left|\sum\right|\le{2|c|\over3}\,\gamma\sum_{m,n}|n|^{1-\varepsilon/2}
(|m|^{\varepsilon/2}+|n-m|^{\varepsilon/2})|m|
\,|\widehat{w}_m||\widehat{w}_{n-m}||\widehat{w}_{-n}|.
\label{tra}\end{equation}
By the Cauchy-Bunyakovsky-Schwarz inequality, for $\varepsilon>0$
\begin{equation}
\left|\sum_n|n|^{1-\varepsilon/2}|\widehat{w}_{-n}|\right|=
\left|\sum_n|n|^{-(1+\varepsilon)/2}|n|^{3/2}|\widehat{w}_{-n}|\right|\le
\left(\sum_n|n|^{-1-\varepsilon}\right)^{\!\!1/2}\|w\|_{3/2}.
\label{tra1}\end{equation}
By the Cauchy-Bunyakovsky-Schwarz and H\"older's inequalities,
for $0\le\varepsilon\le1$
\begin{align}
&\left|\sum_m|m|^{1+\varepsilon/2}|\widehat{w}_m||\widehat{w}_{n-m}|\right|\le
\|w\|_0\|w\|_{1+\varepsilon/2}\nonumber\\
=&\|w\|_0\left(\sum_m(|m|^3|\widehat{w}_m|^2)^{\varepsilon}
|m|^2|\widehat{w}_m|^2)^{1-\varepsilon}\right)^{\!\!1/2}
\le\|w\|_0\|w\|_{3/2}^{\varepsilon}\|w\|_1^{1-\varepsilon}\label{tra2}
\end{align}
and
\begin{equation}
\left|\sum_m|m||n-m|^{\varepsilon/2}|\widehat{w}_m||\widehat{w}_{n-m}|\right|
\le\|w\|_{\varepsilon/2}\|w\|_1.
\label{tra3}\end{equation}
Inequalities \rf{tra}--\rf{tra3} imply
\begin{align*}
\left|\sum\right|\le&{2|c|\over3}\,\gamma\,(\|w\|_0\|w\|_{3/2}^{\varepsilon}
\|w\|_1^{1-\varepsilon}+\|w\|_{\varepsilon/2}\|w\|_1)
\left(\sum_n|n|^{-1-\varepsilon}\right)^{\!\!1/2}\!\!\|w\|_{3/2}\\
\le&Q_5\,(\|w\|_0+\|w\|_{\varepsilon/2})\,\|w\|_1^{1-\varepsilon},
\end{align*}
where~~$\displaystyle
Q_5={2\beta|c|\over3}\,\left(\sum_n|n|^{-1-\varepsilon}\right)^{\!\!1/2}$.

Thus, we find from \rf{enin}:
\begin{align*}
&{1\over2}\,{d\over dt}\left(\|w\|^2_0+a\|w\|^2_1+2\beta(1+\varepsilon)
I(\Z w\Z)\right)\\
\le&-d\|w\|^2_0+Q_5\,(\|w\|_0+\|w\|_{\varepsilon/2})
\|w\|_1^{1-\varepsilon}+\Gn e\Gn_{\beta,\,0}\|w\|_0,\rule{0mm}{4ex}
\end{align*}
whereby
\begin{equation}
{d\zeta\over dt}\le-d\zeta+{Q_5\over a}\,\zeta^{1-\varepsilon}+\Gn e\Gn_{\beta,\,0},
\label{zbou}\end{equation}
where it is denoted
$$\zeta^2=\|w^{(N)}\|^2_0+a\|w^{(N)}\|^2_1+2\beta(1+\varepsilon)I(\Z w^{(N)}\Z).$$

For $d<0$, by Gronwall's lemma $\zeta\le\zeta_0\,\e^{\mu t}$ for any
$\mu>-d$; since also \hbox{$2\beta(1+\varepsilon)I(\Z w^{(N)}\Z)\le\zeta^2$}
and by virtue of \rf{ord}, we have $\Z w^{(N)}\Z\le\zeta_1\,\e^{2\mu t/(1-\varepsilon)}$;
here $\zeta_0$ and $\zeta_1$ are suitable positive constants. Consequently,
$$\Gn\varphi^{(N)}\Gn_{\beta/(1+\zeta_1^{1+\varepsilon}
\e^{(2\mu(1+\varepsilon)/(1-\varepsilon))t}),1}\le
\Gn\varphi^{(N)}\Gn_{\beta/(1+\Z w^{(N)}\Z^{1+\varepsilon}),1}
=\|w^{(N)}\|_1\le{\zeta_0\over\sqrt{a}}\,\e^{\mu t}.$$
This bound is uniform in $N$, and therefore in the limit $N_k\to\infty$ we obtain
$$\Gn\varphi\Gn_{\beta/(1+\zeta_1^{1+\varepsilon}\e^{(2\mu(1+\varepsilon)
/(1-\varepsilon))t}),1}\le{\zeta_0\over\sqrt{a}}\,\e^{\mu t}.$$
Hence the width of the analyticity strip of $\varphi$ around the real axis
is bounded from below by an exponentially decaying quantity,
$\beta/(1+\zeta_1^{1+\varepsilon}\e^{(2\mu(1+\varepsilon)/(1-\varepsilon))t})$.

For $d\ge0$, \rf{zbou} reduces to
$${d\zeta\over dt}\le Q_5\,\zeta^{1-\varepsilon}+\Gn e\Gn_{\beta,\,0}.$$
Integrating this inequality yields $\zeta^{\varepsilon}\le\zeta_2t+\zeta_3$.
Since $2\beta(1+\varepsilon)I(\Z w^{(N)}\Z)\le\zeta^2$, \rf{ord} implies
$\Z w^{(N)}\Z^{1+\varepsilon}
\le\zeta_4t^{2(1+\varepsilon)/(\varepsilon(1-\varepsilon))}+\zeta_5$.
Consequently,
\begin{align*}
\Gn\varphi^{(N)}\Gn_{\beta/(1+\zeta_5+\zeta_4t^{2(1+\varepsilon)/
(\varepsilon(1-\varepsilon))}),1}\le&
\Gn\varphi^{(N)}\Gn_{\beta/(1+\Z w^{(N)}\Z^{1+\varepsilon}),1}\\
=&\|w^{(N)}\|_1\le(\zeta_2t+\zeta_3)^{1/\varepsilon}/\sqrt{a},\rule{0mm}{4ex}
\end{align*}
where all $\zeta_i$ are suitable positive constants.
Since this bound is uniform in $N$, we obtain in the limit $N_k\to\infty$
$$\Gn\varphi\Gn_{\beta/(1+\zeta_5+\zeta_4t^{2(1+\varepsilon)/(\varepsilon
(1-\varepsilon))}),1}\le(\zeta_2t+\zeta_3)^{1/\varepsilon}/\sqrt{a}.$$
Therefore, for $d\ge0$ the width of the analyticity strip of $\varphi$ around
the real axis is bounded from below by the quantity $\beta/(1+\zeta_5
+\zeta_4t^{2(1+\varepsilon)/(\varepsilon(1-\varepsilon))})$, which decays
algebraically. Within the allowed interval $0<\varepsilon<1$, the exponent
$2(1+\varepsilon)/(\varepsilon(1-\varepsilon))$ takes the minimal value
for $\varepsilon=\sqrt{2}-1$. The optimal exponent that we have thus found is
$2(\sqrt{2}+1)^2$. Q.E.D.

\section{Existence and uniqueness of travelling-wave solutions}\label{trwav}

When the forcing has the form $e(x,t)=e(\xi)$ for $\xi=x-\Omega t$,
the RLWE may have travelling-wave solutions such that
$\varphi(x,t)=\varphi(\xi)$. We establish now their existence.

Substituting $\varphi(x,t)=\varphi(\xi)$ into the RLWE we obtain an equation
for the wave profile $\varphi$:
\begin{equation}
-\Omega(\varphi'-a\varphi''')+b\varphi'+c\varphi\varphi'+d\varphi+e(\xi)=0,
\label{trw}\end{equation}
where $'$ denotes henceforth differentiation in $\xi\in\R^1$.
$2\pi$-periodicity in $x$ translates to $2\pi$-periodicity in $\xi$.

{\it Theorem 3.} Suppose $a\Omega d\ne0$. If $e(\xi)\in C^\infty(\R^1)$ is
$2\pi$-periodic, then there exists a $2\pi$-periodic solution to \rf{trw},
$\varphi(\xi)\in C^\infty(\R^1)$, for any constants $a>0$, $b,\,c$ and $d\ne0$.
If the forcing is weak:
\begin{equation}
\|e\|_0<{|d|\over2\sqrt{\sum_{n\ne0}|p_n|^2}},
\label{uni}\end{equation}
where quantities $p_n$ are defined in \rf{pdef} below, the travelling-wave
solution to the RLWE is unique.

{\it Proof}.

$i.$ We seek a solution to \rf{trw} in the form of a Fourier series
$$\varphi(\xi)=\sum_n\widehat{\varphi}_n\,\e^{\i n\xi}.$$
The travelling-wave RLWE then reduces to the system of equations
\begin{equation}
-\i n(1+an^2)\Omega\widehat{\varphi}_n+(d+\i bn)\widehat{\varphi}_n
+{\i cn\over2}\sum_m\widehat{\varphi}_m\widehat{\varphi}_{n-m}+\widehat{e}_n=0.
\label{geqtw}\end{equation}

Equation \rf{geqtw} for $n=0$ (i.e., the average of \rf{trw} over $\xi$)
yields $\widehat{\varphi}_0=-\widehat{e}_0/d$.

Dividing \rf{geqtw} by
$-\i(a\Omega n^3+(\Omega+c\widehat{e}_0/d-b)n+\i d)$, we obtain for $n\ne0$
\begin{equation}
\widehat{\varphi}_n=p_n\sum_{0\ne m\ne n}\widehat{\varphi}_m\widehat{\varphi}_{n-m}+q_n,
\label{Neqtw}\end{equation}
where it is denoted
\begin{align}
p_n=&{cn\over2(a\Omega n^3+(\Omega+c\widehat{e}_0/d-b)n+\i d)},\label{pdef}\\
q_n=&-\,{\i\widehat{e}_n\over a\Omega n^3+(\Omega+c\widehat{e}_0/d-b)n+\i d}. \label{qdef}
\end{align}
The system of equations \rf{Neqtw} does not involve an equation for $n=0$.
To simplify notation, we henceforth formally assume that $\widehat{\varphi}_0=0$
in \rf{Neqtw}.

$ii.$ We have thus rendered the travelling-wave RLWE as a fixed-point problem
$\varphi={\cal A}\varphi$, where the operator $\cal A$ is defined by the r.h.s.
of \rf{Neqtw}:
$${\cal A}:\sum_{n\ne0}\varphi_n(t)\,\e^{\i n\xi}\mapsto\sum_{n\ne0}\left(p_n
\sum_{0\ne m\ne n}\widehat{\varphi}_m\widehat{\varphi}_{n-m}+q_n\right)\e^{\i n\xi}.$$
We will seek a solution in the subspace of zero-mean functions of the Sobolev
space $H^1(\T^1)$ (the norm $\|\cdot\|_2$ in $H^s(\T^1)$ was defined in the
previous section). Existence of solutions to the fixed-point problem
\rf{Neqtw} is guaranteed by the Leray--Schauder principle (\cite{LeSch},
see also \cite{Lad}) under two conditions:

1) Any solution to the equation
\begin{equation}
\varphi=\mu{\cal A}\varphi
\label{mueq}\end{equation}
belongs to a ball
in $H^1(\T^1)$ of a radius independent of $\mu$ for $0\le\mu\le1$.

2) The operator ${\cal A}:\,H^1(\T^1)\to H^1(\T^1)$ is compact, i.e.,
${\cal A}(\varphi^n)$ strongly converges in $H^1(\T^1)$
for any sequence $\varphi^n$, weakly converging in $H^1(\T^1)$.

To establish 1), note that \rf{mueq} is equivalent to the system of equations
$$(-\i n(1+an^2)\Omega+d+\i bn)\widehat{\varphi}_n+{\i\mu cn\over2}
\sum_m\widehat{\varphi}_m\widehat{\varphi}_{n-m}+\mu\widehat{e}_n=0.$$
We multiply this equation for $n\ne0$ by
$\widehat{\varphi}_{-n}$ and sum the results over $n$ to find
$$d\|\varphi\|_0^2=-\mu\sum_n\widehat{\varphi}_n\widehat{e}_n,$$
which implies
$$\|\varphi\|_0\le\mu\|e\|_0/|d|.$$

We multiply now \rf{Neqtw} by $|n|^{2s+2}\widehat{\varphi}_{-n}$;
summation over $n$ then yields
$$\|\varphi\|^2_{s+1}=\mu\sum_{n\ne0}p_n|n|^{2s+2}\sum_{0\ne m\ne n}
\widehat{\varphi}_m\widehat{\varphi}_{n-m}\widehat{\varphi}_{-n}+
\mu\sum_{n\ne0}q_n|n|^{2s+2}\widehat{\varphi}_{-n}.$$
By virtue of the inequality $|n|^{2s}\le R_s(|m|^{2s}+|n-m|^{2s})$, valid
for all $s>0$ and $R_s=\max(2^{2s-1},1)$,
\begin{align*}
&\|\varphi\|^2_{s+1}\le\mu R_{s+1}\!\!\sum_{0\ne m\ne n\ne0}\!\!|p_n||n|^{s+1}(|m|^{s+1}+|n-m|^{s+1})
|\widehat{\varphi}_m||\widehat{\varphi}_{n-m}||\widehat{\varphi}_{-n}|\\
&\hspace*{3em}+\mu\left(\sum_{n\ne0}|q_n|^2|n|^{2s+2}\right)^{\!\!1/2}\!\!\|\varphi\|_{s+1}\\
&\le\mu R_{s+1}\sum_{n\ne0}|p_n||n|^{s+1}|\widehat{\varphi}_{-n}|~
2\|\varphi\|_0\|\varphi\|_{s+1}+\mu\,\left(\sup_{n\ne0}|q_n||n|^3\right)\|e\|_{s-2}\|\varphi\|_{s+1}\\
&\le2\mu R_{s+1}\|\varphi\|_0\|\varphi\|_{s+1}
\|\varphi\|_s\left(\sum_{n\ne0}|p_n|^2|n|^2\right)^{\!\!1/2}
\!\!+\mu\,\left(\sup_{n\ne0}|q_n||n|^3\right)\|e\|_{s-2}\|\varphi\|_{s+1},
\end{align*}
whereby
\begin{equation}
\|\varphi\|_{s+1}\le2\mu R_{s+1}\left(\sum_{n\ne0}|p_n|^2|n|^2\right)^{\!\!1/2}
\|\varphi\|_0\|\varphi\|_s+\mu\,\left(\sup_{n\ne0}|q_n||n|^3\right)\|e\|_{s-2}.
\label{prebo}\end{equation}
Assuming here $s=0$ we find that any solution $\varphi$ to the problem
\rf{mueq} for $0\le\mu\le1$ belongs to the ball
$$\|\varphi\|_1\le\left({2\|e\|_0\over|d|}\right)^2\left(\sum_{n\ne0}|p_n|^2|n|^2
\right)^{\!\!1/2}+\left(\sup_{n\ne0}|q_n||n|^3\right)\|e\|_{-2},$$
as required.

To establish 2), consider a weakly converging sequence $\varphi^k(\xi)$
in $H^1(\T^1)$. By properties of weak convergence, functions
$\varphi^k(\xi)$ are uniformly bounded in $H^1(\T^1)$: $\|\varphi^k\|_1\le A$.
By the Sobolev embedding theorem, weak convergence in $H^1(\T^1)$ implies
strong convergence in $H^1(\T^1)$: for any $\epsilon>0$ and $s<1$ there exists
$K(s)$ such that $\|\varphi^{k'}-\varphi^{k''}\|_s\le\epsilon$
provided $k'>K(s)$ and $k''>K(s)$. We need to show that
$$\|{\cal A}(\varphi^{k'})-{\cal A}(\varphi^{k''})\|_1
=\left\|\sum_{n\ne0}\left(p_n\sum_{0\ne m\ne n}\left(
\widehat{\varphi}^{k'}_m\widehat{\varphi}^{k'}_{n-m}-\widehat{\varphi}^{k''}_m
\widehat{\varphi}^{k''}_{n-m}\right)\right)\e^{\i n\xi}\right\|_1\to0$$
for $k',k''\to\infty$. In terms of $\theta^{\,k',k''}=\varphi^{\,k'}-\varphi^{\,k''}$ and
$\widehat{\theta}^{\,k',k''}=\widehat{\varphi}^{\,k'}-\widehat{\varphi}^{\,k''}$,
\begin{align*}
&\|{\cal A}(\varphi^{k'})-{\cal A}(\varphi^{k''})\|^2_1\\
=&\!\sum_{n\ne0}\!\left(|p_n|^2|n|^2\!\sum_{0\ne m\ne n}
\!(\widehat{\varphi}^{\,k'}_m\,\widehat{\theta}^{\,k',k''}_{n-m}
+\widehat{\varphi}^{\,k''}_{n-m}\,\widehat{\theta}^{\,k',k''}_m)
\!\!\sum_{0\ne l\ne-n}\!(\widehat{\varphi}^{\,k'}_l\,\widehat{\theta}^{\,k',k''}_{-n-l}
+\widehat{\varphi}^{\,k''}_{-n-l}\,\widehat{\theta}^{\,k',k''}_l)\!\right)\\
\le&\sum_{n\ne0}|p_n|^2|n|^2\,(\|\varphi^{k''}\|_0+\|\varphi^{k'}\|_0)^2
\|\theta^{\,k',k''}\|^2_0
\le4A^2\left(\sum_{n\ne0}|p_n|^2|n|^2\right)\|\theta^{\,k',k''}\|^2_0.
\end{align*}
This proves the required strong convergence of ${\cal A}(\varphi^k)$ in
$H^1(\T^1)$ for $k\to\infty$.

$iii.$ Solutions $\varphi$ to the travelling-wave RLWE have finite norms
in any Sobolev space $H^s(\T^1)$ and hence are infinitely differentiable.
This follows directly from inequality \rf{prebo} for $\mu=1$
in combination with induction in integer $s>0$.

$iv.$ The number of solutions to the travelling-wave RLWE for given
parameter values is unknown, unless the coefficients $p_n$ and/or the energy
$\|\varphi\|_0$ are small, in which case the solution is unique.

Suppose there exist two solutions $\varphi'$ and $\varphi''$. We denote
$\theta=\varphi'-\varphi''$ and
$\widehat{\theta}=\widehat{\varphi}'-\widehat{\varphi}''$ and find
\begin{align*}
\|\theta\|^2_0&=\|{\cal A}(\varphi')-{\cal A}(\varphi'')\|^2_0\\
&=\sum_{n\ne0}\left(|p_n|^2\sum_{0\ne m\ne n}
(\widehat{\varphi}\,'_m\,\widehat{\theta}_{n-m}
+\widehat{\varphi}\,''_{n-m}\,\widehat{\theta}_m)
\sum_{0\ne l\ne-n}(\widehat{\varphi}\,'_l\,\widehat{\theta}_{-n-l}
+\widehat{\varphi}\,''_{-n-l}\,\widehat{\theta}_l)\right)\\
&\le\sum_{n\ne0}|p_n|^2\,(\|\varphi'\|_0+\|\varphi''\|_0)^2\|\theta\|^2_0.\rule{0mm}{4ex}
\end{align*}
Thus coexistence of distinct solutions satisfying
$$\|\varphi\|_0<\left(4\sum_{n\ne0}|p_n|^2\right)^{-1/2}$$
is ruled out. Since any solution to the travelling-wave RLWE has a bounded norm
$\|\varphi\|_0\le\|e\|_0/|d|$, the problem has a unique solution provided
inequality \rf{uni} holds true. Q.E.D.

\section{Non-well-posedness of the non-damped travelling-wave RLWE}\label{nonpos}

A problem of physical relevance is said, following Hadamard, to be well-posed,
if it possesses a solution that is unique and depends continuously on the data.
The evolutionary problem for the non-damped ($d=0$) RLWE (the BBM equation)
is well-posed \cite{BBM}. Instead of developing the existence theory
for travelling waves for $d=0$, we show here that, by contrast,
the travelling-wave problem for the RLWE \rf{trw} for $d=0$ is not
well-posed, since arbitrarily large solutions can exist for a forcing
of whichever small amplitude.

We consider fast oscillating (both in space and time)
solutions to \rf{trw} of the form
\begin{equation}
\varphi(\xi)=\Omega^\beta\Phi(\eta),\qquad\eta\equiv\Omega^\alpha\xi
\label{wanz}\end{equation}
in the limit $\Omega\to\infty$. Substituting the ansatz \rf{wanz}
into \rf{trw} yields
\begin{equation}
-\Omega^{1+\alpha+\beta}\Phi'+a\Omega^{1+3\alpha+\beta}\Phi'''
+b\Omega^{\alpha+\beta}\Phi'+c\Omega^{\alpha+2\beta}\Phi\Phi'+e(\eta)=0,
\label{Otrw}\end{equation}
the prime $'$ denoting in this section differentiation in the fast variable
$\eta$. In this section we assume $a>0$ and $c<0$ (the important condition here
is $c\ne 0$; the convention about the sign of $c$ is technical since \rf{Otrw}
has the symmetry $\Phi\to-\Phi$, $c\to-c$). Note that for such coefficients
we might reduce equation \rf{Otrw} to the one for $a=b=c=1$ (provided $b\ne0$)
by appropriately rescaling $\Omega$, $\eta$ and $\Phi$.

The two terms defining the nature of the problem, i.e., the ones involving
the third-order derivative and the nonlinearity, balance each other if
$\beta=1+2\alpha$. If $\alpha$ and $\beta$ are rational, it is natural to seek
$\Phi$ in the form of power series in $\Omega^{-1}$ in an appropriate
fractional power. The simplest case is $\alpha=1$, $\beta=3$.
For these parameter values we consider the series
\begin{equation}
\Phi(\eta)=\sum_{i\ge0}\Omega^{-i}\Phi_i(\eta).
\label{ePhi}\end{equation}
Substituting \rf{ePhi} into \rf{Otrw}, expanding and collecting all the terms
involving $\Omega^{7-i}$ for some $i\ge0$, we obtain a hierarchy of equations
for $\Phi_i(\eta)$:
\begin{equation}
-\Phi'_{i-2}+a\Phi'''_i+b\Phi'_{i-3}+c\sum_{j=0}^i\Phi_j\Phi'_{i-j}=0
\label{Phii}\end{equation}
(assuming that the amplitude of the forcing $e(\eta)$ is so small that it does
not contribute to \rf{Phii} at this level).

\begin{figure}
\centerline{\includegraphics[width=6cm]{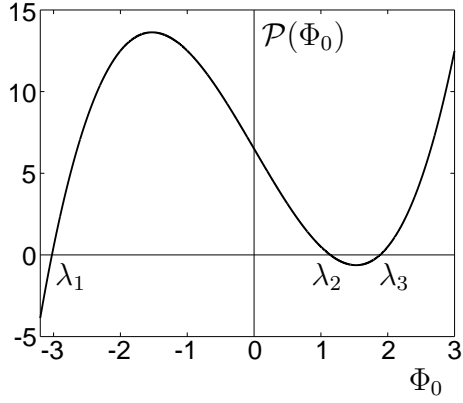}}
\vspace*{-47mm}\hspace*{73mm}${\cal P}(\Phi_0)$\\
\rule{0mm}{30.5mm}\hspace*{45.5mm}$\lambda_1$\hspace*{3cm}$\lambda_2$
\hspace*{3.5mm}$\lambda_3$\\
\rule{0mm}{13mm}\hspace*{92.5mm}$\Phi_0$

\caption{A plot of the cubic polynomial ${\cal P}(\Phi_0)$ in the r.h.s.~of
\rf{Phi01} for $Q=3$, $C_0=6.5$, $C_1=-7$.}
\label{fig1}\end{figure}

$i$. For $i=0$, \rf{Phii} reduces to
\begin{equation}
\Phi'''_0-Q\Phi_0\Phi'_0=0,
\label{Phi03}\end{equation}
where it is denoted $Q=-c/a$ (by our convention $Q>0$). Integrating \rf{Phi03}
in~$\eta$ once, we find
\begin{equation}
\Phi''_0={1\over2}\,(Q\Phi^2_0+C_1).
\label{Phi02}\end{equation}
Using the standard techniques, we reduce the order of \rf{Phi02}
by regarding $\Phi_0$ as a new independent variable and $\Phi'_0$ as an unknown
function of this variable; integrating \rf{Phi02} in $\Phi_0$ then yields
\begin{equation}
(\Phi'_0)^2={\cal P}(\Phi_0)\equiv{Q\over3}\,\Phi^3_0+C_1\Phi_0+C_0,
\label{Phi01}\end{equation}
where $C_1$ and $C_0$ are some constants. We assume henceforth
$4C_1^3<-9QC^2_0$, whereby the polynomial ${\cal P}(\Phi_0)$ in the r.h.s.~of
\rf{Phi01} has three distinct real roots $\lambda_1<\lambda_2<\lambda_3$
(see a sketch of the plot of ${\cal P}(\Phi_0)$ in Fig.~\ref{fig1}).

Solutions to \rf{Phi01} can be expressed in terms of the Weierstrass elliptic
function (see \cite{Akh,Abr})
$$\wP(z;\omega_1,\omega_2)={1\over z^2}+
\sum_{|n_1|+|n_2|\ne0}\left({1\over(z+2n_1\omega_1+2n_2\omega_2)^2}
-{1\over(2n_1\omega_1+2n_2\omega_2)^2}\right)$$
that is holomorphic and double-periodic, the periods being $2\omega_1$
and $2\omega_2$, and solves equations
\begin{equation}
\left({d\over dz}\wP\right)^2=4\wP^3-g_2\wP-g_3
\label{WODE}\end{equation}
and
\begin{equation}
z=\int_{\wp(z;\omega_1,\omega_2)}^\infty(4\Phi^3-g_2\Phi-g_3)^{-1/2}d\Phi
\label{inve}\end{equation}
on the complex plane $z\in\C^1$. The Weierstrass elliptic function was employed
to solve a generalised BBM equation in \cite{Nick}. The rescaled function
$\widetilde\Phi=(Q/12)\Phi_0$ satisfies ODE \rf{WODE} for $g_2=-QC_1/12$ and
$g_3=-Q^2C_0/144$. The half-periods $\omega_i$ can be found from the conditions
$$g_2=60\sum_{|n_1|+|n_2|\ne0}(2n_1\omega_1+2n_2\omega_2)^{-4},\qquad
g_3=140\sum_{|n_1|+|n_2|\ne0}(2n_1\omega_1+2n_2\omega_2)^{-6}.$$
Since the roots of the r.h.s.~of \rf{WODE}, $e_i=(Q/12)\lambda_i$, are real,
one of the half-periods (say, $\omega_1$) is real, and the other one
(respectively, $\omega_2$) is imaginary.
Separating variables in \rf{WODE} and taking into account \rf{inve}, we find
\begin{equation}
\widetilde\Phi(\eta)=\wP\left(\int_{\widetilde\Phi(\eta_0)}^\infty(4\Phi^3
-g_2\Phi-g_3)^{-1/2}d\Phi+\eta_0-\eta;\,\omega_1,\omega_2\right).
\label{path}\end{equation}
The three quantities $\wP(\omega_1;\omega_1,\omega_2)$,
$\wP(\omega_2;\omega_1,\omega_2)$ and $\wP(\omega_1+\omega_2;\omega_1,\omega_2)$
coincide with the roots $e_i$, and hence, by \rf{inve},
$$\omega_1=\int_{e_3}^\infty(4\Phi^3-g_2\Phi-g_3)^{-1/2}d\Phi,\qquad
\omega_2=\int_{e_2}^{e_3}(4\Phi^3-g_2\Phi-g_3)^{-1/2}d\Phi.$$
This removes the ambiguity in the choice of branches of the square root
in the path of integration in the r.h.s.~of \rf{path}. Using the addition
formula for the Weierstrass elliptic function and the relations
$\wP(\omega_1+\omega_2;\omega_1,\omega_2)=e_2$,
$\wP'(\omega_1+\omega_2;\omega_1,\omega_2)=0$, we obtain from
$\rf{path}$ the solution in the form that does not involve complex numbers:
$$\widetilde\Phi(\eta)=e_2+{(e_2-e_1)(e_2-e_3)\over\wP\left(
\displaystyle\int_{\widetilde\Phi(\eta_0)}^{e_2}(4\Phi^3
-g_2\Phi-g_3)^{-1/2}d\Phi+\eta_0-\eta;\,\omega_1,\omega_2\right)-e_2}.$$

Furthermore, we can represent the solution in terms of the Jacobi elliptic
func\-tions of modulus $k=\sqrt{(e_2-e_1)/(e_3-e_1)}$ using the identities
(see \cite{Akh,Abr})
$$\wP\!\left(\!{q\over\sqrt{e_3-e_1}};\,\omega_1,\omega_2\!\right)\!=e_1+{e_3-e_1\over{\rm sn}^2(q)}
=e_2+(e_3-e_1){{\rm dn}^2(q)\over{\rm sn}^2(q)}
=e_3+(e_3-e_1){{\rm cn}^2(q)\over{\rm sn}^2(q)}.$$

However, rather than applying the above results of the theory
of elliptic functions, it appears more instructive to establish the properties
of the solution that are important for our purposes
by directly inspecting \rf{Phi03}--\rf{Phi01}. Consider a solution to the ODE
\rf{Phi01} such that $\lambda_1<\Phi_0(0)<\lambda_2$. To be specific, let
$\Phi'_0(0)$ satisfying \rf{Phi01} be positive. Thus, on increasing $\eta$,
$\Phi_0$ is growing till it approaches the value $\lambda_2$. The ODE
\rf{Phi01} can be expressed as
$$\Phi'_0=C(\Phi_0)\sqrt{\lambda_2-\Phi_0}.$$
For $\Phi_0\approx\lambda_2$,
$C(\Phi_0)\approx\sqrt{Q(\lambda_2-\!\lambda_1)(\lambda_3-\!\lambda_2)/3}$
is a smooth function bounded from below by a positive constant. Consequently,
$\Phi_0$ takes the limit value $\lambda_2$ at a finite $\eta=\eta_0$.
For $\Phi_0=\lambda_2$, the r.h.s.~of \rf{Phi02} is non-zero, and hence
at $\eta=\eta_0$ the sign of $\Phi'_0$ changes and $\Phi_0$ begins to decrease.
Separation of variables in \rf{Phi01} yields
$$\pm\int^{\lambda_2}_{\Phi_0(\eta)}\left({Q\over3}\,\Phi^3+C_1\Phi+C_0
\right)^{-1/2}d\Phi=\eta-\eta_0,$$
where the sign in the l.h.s. is `$-$' for $\eta<\eta_0$ and `$+$'
for $\eta>\eta_0$. By virtue of this relation, $\Phi_0(\eta)$ is a symmetric
function of $\eta$ about $\eta_0$: $\Phi_0(\eta_0+\eta)=\Phi_0(\eta_0-\eta)$.

\begin{figure}
\centerline{\includegraphics[width=6cm]{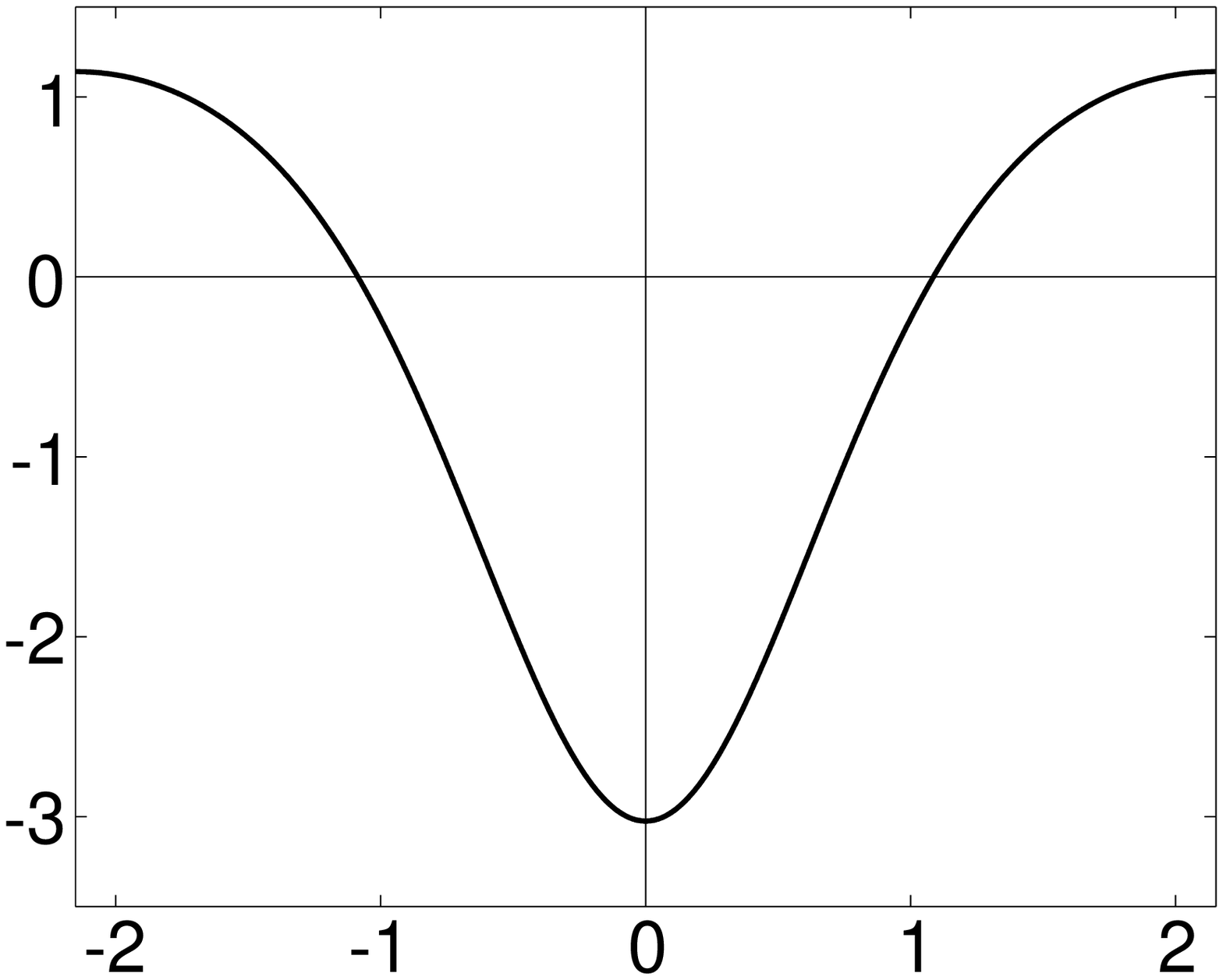}\hspace*{1cm}
\includegraphics[width=6cm]{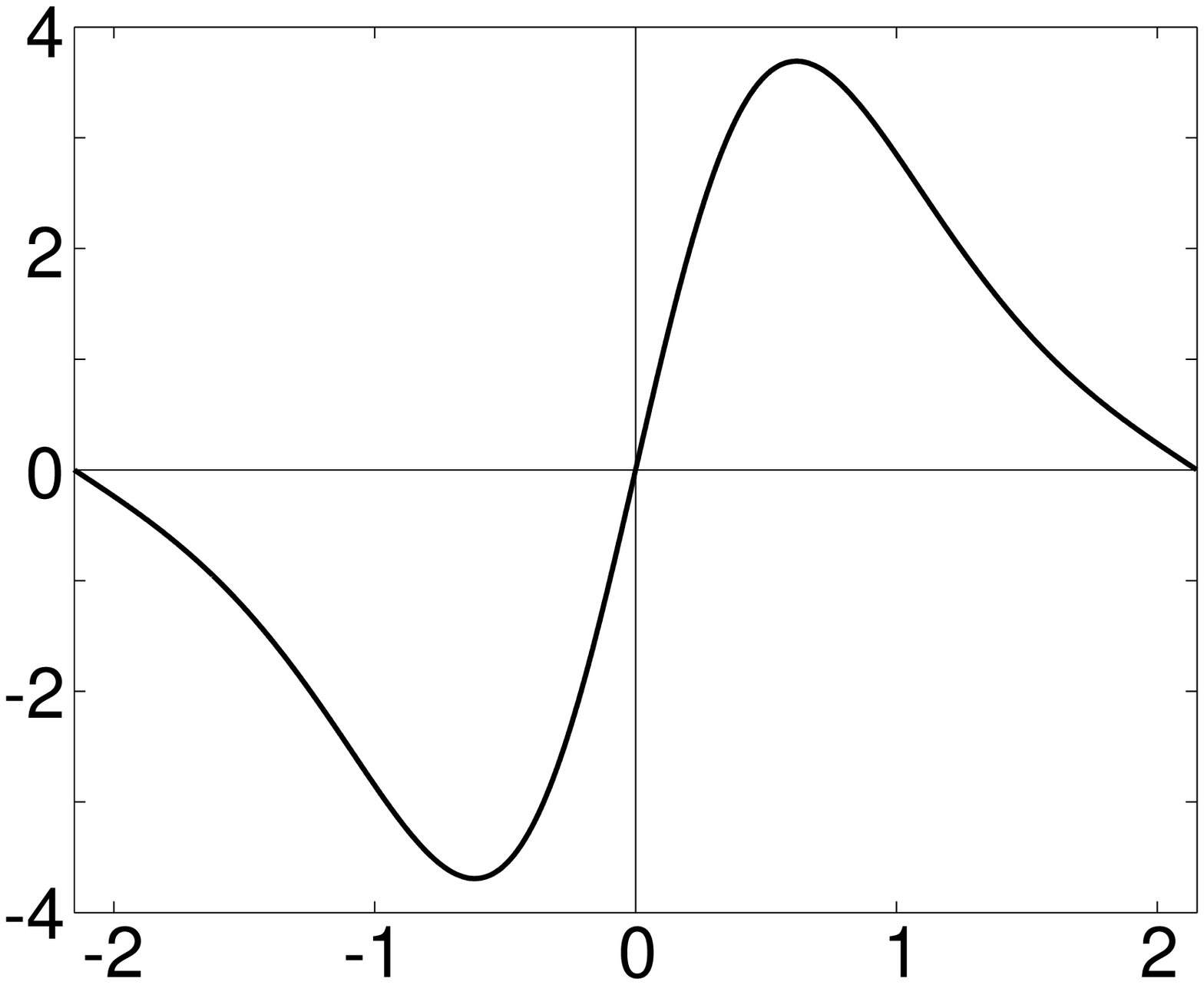}}
\vspace*{-48mm}\hspace*{24.5mm}$\Phi_0(\eta)$\hspace*{61.5mm}$\Phi'_0(\eta)$\\
\rule{0mm}{10.5mm}\hspace*{6cm}$\eta$\\
\rule{0mm}{7.5mm}\hspace*{131.5mm}$\eta$\\~\\~\\~\\

\caption{Plots of a sample solution $\Phi_0(\eta)$ to \rf{Phi01} for $Q=3$,
$C_0=6.5$, $C_1=-7$ (left panel) and its derivative $\Phi'_0(\eta)$ (right
panel).}
\label{fig2}\end{figure}

By a similar argument, $\Phi_0$ continues to decrease till
$\Phi_0(\eta_1)=\lambda_1$ for some $\eta=\eta_1$, and subsequently
the process repeats itself: there exists an infinite sequence $\eta_k$
such that $\Phi_0(\eta_{2k})=\lambda_2$ and
$\Phi_0(\eta_{2k+1})=\lambda_1$. Moreover, $\Phi_0(\eta)$ is symmetric
in $\eta$ about each $\eta_k$. Thus, $\Phi_0(\eta)$ is periodic in $\eta$,
with the half-period $E/2=\eta_{k+1}-\eta_{k}$ (this value being independent
of $k$). In what follows we fix the origin of the variable $\eta$ by letting
$\eta_1=0$. Plots of a sample solution $\Phi_0(\eta)$ to \rf{Phi01}--\rf{Phi03}
computed for $Q=3$, $C_0=6.5$, $C_1=-7$ and its derivative $\Phi'_0(\eta)$
are shown in Fig.~\ref{fig2}.

$ii$. For $i>0$, \rf{Phii} becomes
\begin{equation}
a{\cal L}\Phi_i-\Phi'_{i-2}+b\Phi'_{i-3}+c\sum_{j=1}^{i-1}\Phi_j\Phi'_{i-j}=0,
\label{Phii2}\end{equation}
for $i=1$ reducing to
\begin{equation}
{\cal L}\Phi_1=0.
\label{term1}\end{equation}
Here $\cal L$ is the operator
of linearisation of \rf{Phi03} in the vicinity of $\Phi_0(\eta)$:
$${\cal L}f=f'''_i-Q(\Phi_0 f)'.$$
It is assumed to act in the Lebesgue space of zero-mean functions that have
the same period $E$ in $\eta$ as $\Phi_0$. The adjoint operator is
$${\cal L}^*f=-f'''_i+Q\{\Phi_0f'\},$$
where
$$\langle f\rangle={1\over E}\,\int_{-E/2}^{E/2}f(\eta)\,d\eta\qquad
\mbox{and}\qquad\{f\}=f-\langle f\rangle$$
denote the average of function $f$ over the period $E$ of $\Phi_0$
and its oscillatory part, respectively. Evidently, operators $\cal L$ and
${\cal L}^*$ map the subspace of even functions (i.e., such that
$f(\eta)=f(-\eta)$\,), into the subspace of odd functions (i.e.,
$f(\eta)=-f(-\eta)$\,), and vice versa.

In order to determine the solvability conditions for equations \rf{Phii2}, we need
to examine the kernel of ${\cal L}^*$.
By \rf{Phi03}, ${\cal L}^*\{\Phi_0\}=0$. Differentiating \rf{Phi03}
in $\eta$ yields ${\cal L}\Phi'_0=0$ (this is a manifestation of translation
invariance of equations \rf{Phi01}--\rf{Phi03}\,). Thus, the kernels of $\cal L$
and ${\cal L}^*$ are at least one-dimensional. Actually, generically
$\dim\ker{\cal L}=\dim\ker{\cal L}^*=2$, the kernels involving generalised
eigenfunctions and the operators having $2\times2$ Jordan cells associated
with the eigenvalue~0. To see this, consider solutions to the problems
$$S''_\nu-Q(\Phi_0S_\nu-\nu)=0,\qquad S_\nu(0)=1,\qquad S'_\nu(0)=0$$
and the linear combination $S(\eta)=\mu S_0(\eta)+(1-\mu)S_1(\eta)$, where $\mu$
is found from the condition $S'(E/2)=0$. As we have established, $\Phi_0$
is symmetric about the points $kE/2$, where $k$ is integer. Using this,
it is easy to show that $S(\eta)$ is also symmetric about these points, and
thus is $E$-periodic. By construction, $S(\eta)$ satisfies the equations
\begin{equation}
S'''_i-Q(\Phi_0S)'=0\qquad\Leftrightarrow\qquad{\cal L}\{S\}=Q\langle S\rangle\Phi'_0.
\label{kerL}\end{equation}
Thus, ${\cal L}\{S\}\ne0$ unless
$\langle S\rangle=0$, but ${\cal L}^2\{S\}=0$, i.e., $\{S\}$ is a generalised
eigenfunction associated with the eigenvalue 0 (clearly, $\Phi'_0$ and $\{S\}$
are linearly independent: the former eigenfunction is odd while the latter is
even). The respective odd generalised eigenfunction from the kernel
of ${\cal L}^*$ is $\left\{\int_0^\eta\{S\}\,d\eta\right\}$.

\begin{figure}
\centerline{\includegraphics[width=6cm]{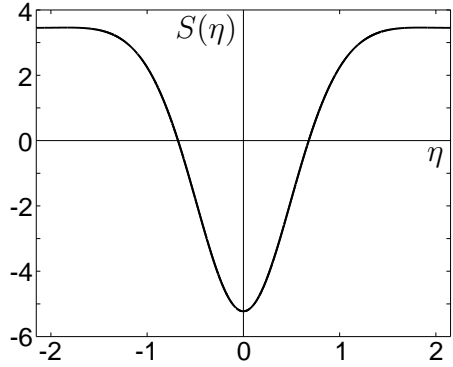}}
\vspace*{-48mm}\hspace*{62mm}$S(\eta)$\\
\rule{0mm}{15mm}\hspace*{95.5mm}$\eta$\\~\\~\\~\\~\\

\caption{A plot of the function $S$ computed
for a sample solution $\Phi_0$ to \rf{Phi01} shown in Fig.~\ref{fig2}.}
\label{fig3}\end{figure}

We present in Fig.~\ref{fig3} a plot of the function $S$ that was computed
for a sample solution to \rf{Phi03}--\rf{Phi01} $\Phi_0$ shown
in Fig.~\ref{fig2}. $\langle S\rangle=0.93314$ is non-zero
beyond numerical accuracy (the Lebesgue norm of $S$ is 6.54875).
We have checked numerically that the kernel of $\cal L$ is two-dimensional.

The theorem on the Fredholm alternative implies that an equation of the form
\begin{equation}
{\cal L}f=u
\label{Lfu}\end{equation}
is solvable in the space of zero-mean $E$-periodic functions whenever
\begin{equation}
\int_{-E/2}^{E/2}u(\eta)\{\Phi_0\}\,d\eta=0,
\label{cLfu}\end{equation}
and then the solution can be found up to an arbitrary additive term
$\kappa\Phi'_0$. (More precisely, the Fredholm alternative theorem is stated
for linear problems where the operator is a sum of the identity operator and
a compact one \cite{kofo,lyust}; however, by considering \rf{cLfu}
in the Fourier space, it is simple to show that after integrating the equation
three times, we obtain a problem equivalent to \rf{cLfu}, for which the
theorem on Fredholm alternative is readily applicable.)
In particular, the problem \rf{term1} has a general solution
$\Phi_1=\kappa_1\Phi'_0$, where $\kappa_1$ is an arbitrary constant.

$iii$. Thus, \rf{Phii2} specifies $\Phi_i$ up to an arbitrary additive term
$\kappa_i\Phi'_0$. In principle, one starts solving \rf{Phii2} for a given
$i>1$ by satisfying the solvability condition \rf{cLfu} and determining from it
the coefficient $\kappa_j$ for an appropriate $j<i$. However, we can
just set all $\kappa_i=0$. Then all functions $\Phi_i$ are even, equations
\rf{Phii2} have odd non-homogeneous parts, and, $\{\Phi_0\}$ being even,
the solvability conditions \rf{cLfu} are trivially satisfied.
In particular, $\Phi_1=0$ and $\Phi_2=\{S\}/(Q\langle S\rangle)$.

We have therefore shown that one can recursively solve equations \rf{Phii2}
in all orders and determine all terms in the power series \rf{ePhi}.
By construction, a truncated series \rf{ePhi}
$$\Phi_I(\eta)=\sum_{i=0}^I\Omega^{-i}\Phi_i(\eta)$$
is a solution to \rf{Otrw} for the forcing
$$e(\eta)=\sum_{i=\min(7-2I,4-I)}^{6-I}\Omega^ie_i(\eta)$$
(where all $e_i(\eta)$ are of the order of unity). Thus, we have found
an oscillatory solution to the original RLWE \rf{trw} for waves
for $d=0$, whose amplitude grows as O($\Omega^3$), despite it is sustained
by the forcing O($\Omega^{6-I}$) which, for large $I$ and $\Omega$ can be
made arbitrarily small with any fixed number of derivatives. This shows
that the undamped RLWE for waves gives rise to a problem that is not well-posed.

Several remarks are in order. Our construction is not applicable for $d\ne0$
technically because the damping term breaks the symmetry of the solution,
and we cannot argue any more that the solvability
conditions are automatically satisfied. One might try to overcome this
by employing the general procedure, whereby one reintroduces the terms
$\kappa_i\Phi'_0$ from the kernel of the operator of linearisation $\cal L$
into $\Phi_i$ for $i>0$ and satisfies the solvability conditions by solving
the respective equations in $\kappa_i$. However, the system
of equations obtained from the solvability conditions does not have a solution.
The reason for this failure lies in the fact that while we are constructing
a family of solutions to the travelling-wave RLWE that are supposed to grow
with $\Omega$ unboundedly as $\Omega^\beta$, any travelling-wave solution
to the RLWE for $d\ne0$ has a bounded norm $\|\varphi\|_0\le\|e\|_0/|d|$.

The family of travelling waves that we have constructed for $d=0$
is non-unique: asymptotic solutions can be obtained for any
$\alpha>0$, $\beta=1+2\alpha$ with the leading-order term $\Phi_0$ satisfying
equations \rf{Phi03}--\rf{Phi01}. A similar analysis can also be attempted
for $\alpha\le0$, but in this case the equation for the leading term
in the expansion of $\Phi$ differs from \rf{Phi03}.

\section{Asymptotic expansion for a weak forcing}\label{wfor}

We consider now the travelling-wave RLWE \rf{trw} for the forcing proportional
to a small parameter $\epsilon$, i.e., we assume in this section that the term
$e(\xi)$ in \rf{trw} is changed to $\epsilon e(\xi)$. In this case a solution
to \rf{trw} can be sought as an asymptotic power series
\begin{equation}
\varphi(\xi)=\sum_{k>0}\varphi^{(k)}(\xi)\epsilon^k.
\label{asy}\end{equation}
Substituting the series into \rf{trw}, we obtain a transport system of equations
\begin{align}
{\cal M}\varphi^{(1)}=&-e(\xi);\label{o1}\\
{\cal M}\varphi^{(k)}=&-{c\over2}\,{d\over d\xi}\sum_{l=1}^{k-1}
\varphi^{(l)}\varphi^{(k-l)},~~k>1.\label{ok}
\end{align}
Here $\cal M$ denotes the operator
${\cal M}:\varphi\mapsto-\Omega(\varphi'-a\varphi''')+b\varphi'+d\varphi,$
where $'$ denotes the derivative in~$\xi$. Existence of solutions to these
problems follows from Theorem 3 applied for $c=0$.

In terms of the Fourier coefficients of $\varphi^{(k)}$ these equations
take the form, respectively,
\begin{align}
\widehat{\varphi}^{(1)}_n=&q_n;\label{Fo1}\\
\widehat{\varphi}^{(k)}_n=&p_n\sum_m
\sum_{l=1}^{k-1}\widehat{\varphi}^{(l)}_m\widehat{\varphi}^{(k-l)}_{n-m},~~~k>1,
\label{Fok}\end{align}
where
$$p_n={cn\over2(a\Omega n^3+(\Omega-b)n+\i d)},\qquad
q_n=-\,{\i\widehat{e}_n\over a\Omega n^3+(\Omega-b)n+\i d}.$$
Unlike in the previous section, now we do not single out the equation for $n=0$,
since that would imply an undesirable dependence of $p_n$ and $q_n$
on $\epsilon$, as in \rf{pdef}--\rf{qdef}. Note that
$\widehat{\varphi}^{(k)}_0=0$ for $k>1$.

These relations imply
$$\widehat{\varphi}^{(k)}_n=\sum_{\subs{m_1,...,m_k}{m_1+...+m_k=n}}
\zeta_{m_1,...,m_k}q_{m_1}...q_{m_k}.$$
By \rf{Fo1}, for $k=1$ just a single term for $m_1=n$ is present in this sum,
which is $\zeta_n=1$ for any $n$. By \rf{Fok}, the recurrence relation
\begin{align*}
\left.\zeta_{m_1,...,m_{k+1}}=p_{m_1+...+m_{k+1}}\right(&
\zeta_{m_1}\zeta_{m_2,...,m_{k+1}}+...
+\zeta_{m_1,...,m_l}\zeta_{m_{l+1},...,m_{k+1}}\\
&\left.+...+\zeta_{m_1,...,m_k}\zeta_{m_{k+1}}\right)
\end{align*}
holds (there are $k$ terms in the sum in parenthesis here).

\pagebreak
{\it Theorem 4}. Power series \rf{asy} is an asymptotic expansion in $\epsilon$
of the solution $\varphi(\xi)$ to the travelling-wave RLWE.

{\it Proof}. For $K>1$, the residual
$$\theta(\xi)=\varphi(\xi)-\sum_{k=1}^{K-1}\varphi^{(k)}(\xi)\epsilon^k$$
satisfies the equation
\begin{equation}
{\cal M}\theta=-{c\over2}\,{d\over d\xi}\sum_{k=K}^{2K-2}\left(\epsilon^k\,
\sum_{l=1}^{k-1}\varphi^{(l)}\varphi^{(k-l)}\right).
\label{tht}\end{equation}

Multiplying \rf{o1} by $\varphi^{(1)}$, we find $\|\varphi^{(1)}\|_0\le\|e\|_0/|d|$.
Multiplying \rf{Fok} by $\widehat{\varphi}^{(k)}_{-n}$ and summing over $n\ne0$,
we obtain
\begin{align*}
\|\varphi^{(k)}\|_0^2&=\sum_{l=1}^{k-1}\sum_{n\ne0}p_n\widehat{\varphi}^{(k)}_{-n}
\left(\sum_m\widehat{\varphi}^{(l)}_m\widehat{\varphi}^{(k-l)}_{n-m}\right)\\
&\le\sum_{l=1}^{k-1}\left(\sum_{n\ne0}|p_n|^2\right)^{\!\!1/2}\|\varphi^{(k)}\|_0
\|\varphi^{(l)}\|_0\|\varphi^{(k-l)}\|_0,
\end{align*}
whereby
$$\|\varphi^{(k)}\|_0\le\sum_{l=1}^{k-1}\left(\sum_{n\ne0}|p_n|^2\right)^{\!\!1/2}
\|\varphi^{(l)}\|_0\|\varphi^{(k-l)}\|_0.$$
This establishes (using induction in $k$) that all $\varphi^{(k)}$ have
finite norms $\|\varphi^{(k)}\|_0$.

In the Fourier space, equation \rf{tht} in
$$\theta(\xi)=\sum_n\widehat{\theta}_n\,\e^{\i n\xi}$$
takes the form
$$\widehat{\theta}_n=p_n\sum_{k=K}^{2K-2}\epsilon^k\left(
\sum_{l=1}^{k-1}\sum_m
\widehat{\varphi}^{(l)}_m\widehat{\varphi}^{(k-l)}_{n-m}\right).$$
Multiplying it by $\widehat{\theta}_{-n}$ and summing over $n\ne0$, we find
\begin{align*}
\|\theta\|_0^2=&\sum_{k=K}^{2K-2}\left(\epsilon^k\,
\sum_{l=1}^{k-1}\sum_{n\ne0}p_n\widehat{\theta}_{-n}
\left(\sum_m\widehat{\varphi}^{(l)}_m\widehat{\varphi}^{(k-l)}_{n-m}\right)\right)\\
\le&\sum_{k=K}^{2K-2}\left(\epsilon^k\,\sum_{l=1}^{k-1}\left(\sum_{n\ne0}
|p_n|^2\right)^{\!\!1/2}\|\theta\|_0\|\varphi^{(l)}\|_0\|\varphi^{(k-l)}\|_0\right),
\end{align*}
and hence
$$\|\theta\|_0\le\sum_{k=K}^{2K-2}\left(\epsilon^k\,\sum_{l=1}^{k-1}
\left(\sum_{n\ne0}|p_n|^2\right)^{\!\!1/2}\|\varphi^{(l)}\|_0\|\varphi^{(k-l)}\|_0
\right)={\rm O}(\epsilon^K).$$
Q.E.D.

\section{Concluding remarks}

We have presented mathematical results concerning existence, uniqueness,
spatial analyticity and well-posedness of space-periodic evolutionary and
tra\-velling-wave solutions to the RLWE with forcing and damping. This work
has been necessitated by the ongoing intensive numerical study of various
regimes exhibited by solutions to this equation
\cite{He,HeChi03,HeChi04,HeChi05,HeSa89,RemChi07,ReMiChi09,ToChiRem13}.

The techniques used here to analyse the RLWE can also be applied
to the closely related Korteweg-de Vries equation. Well-posedness of problems
that can be stated for this equation is still a topic of active investigation.
Under the condition of spatial periodicity, the Cauchy problem for the KdV
equation was recently proved to be locally well-posed in a class of analytic
functions that can be extended holomorphically in a symmetric strip
of the complex plane around the $x$-axis \cite{himonas}. While we have proved
(section \ref{anal}) that the width of the analyticity strip decays at most
polynomially, it was shown in \cite{himonas} that the uniform radius
of spatial analyticity of solutions to the KdV equation does not shrink as time
goes by.

In the limit of high wave speed, power series expansions of travelling-wave
solutions to the RLWE and the KdV equation differ only in minor details. Thus,
upon introduction of the necessary but non-essential modifications
(in particular, $\beta=2\alpha$ is now required in the ansatz \rf{wanz},
the simplest case being $\alpha=1$, $\beta=2$),
our construction (see section \ref{nonpos}) establishes the lack of continuity
of space-periodic travelling-wave solutions to the KdV equation with respect to
small-amplitude forcing. (Other exact travelling-wave solutions to the KdV
equation with external forcing were recently derived, that involve Jacobi
elliptic functions \cite{Salas,Kudr}; see also \cite{gandarias}.)
% in the absence of forcing, the equation is integrable in \cite{gardner}).

The following problems remain open: Does no shrinking of the width
of the analyticity strip occur for solutions to the RLWE as this happens for
space-periodic solutions to the KdV equation? We have not proved convergence
of the asymptotic power series that we have constructed for travelling-wave
solutions in sections \ref{anal} and \ref{wfor}; do they converge?
Are evolutionary solutions to the RLWE analytic in time? For the sake
of completeness, one would like to extend our results on existence
of travelling waves to cover the case of the absence of damping ($d=0$).
Finally, we have not studied stability of our travelling-wave solutions
to short- or large-scale perturbations; while the former problem can be
addressed numerically, the latter one can be tackled by using the homogenisation
methods similar to those employed in the study of the large-scale magnetic field
generation \cite{Zhdyn}.

Another extension of our work would be an investigation of the shallow-water
wave equation proposed by Camassa and Holm \cite{camassa93}, which could
be modified by adding linear damping and external forcing.

\section*{Acknowledgments}

ACLC and ELR thank the support of CNPq (Brazil) and FAPESP (Brazil). ACLC is
grateful for the award of a Marie Curie International Incoming Fellowship and
the hospitality of Paris Observatory (France). RC, OP and VZ were financed in part
by the grant 11-05-00167-a from the Russian foundation for basic research.
The two-month visits of OP and VZ to the Institute of Aeronautical
Technology (Brazil) were supported by FAPESP (Brazil).

\end{document}